\documentclass[twocolumn]{aastex6}

\newcommand{\dekok}{2013A&A...554A..82D}

\graphicspath{{./figures/}}
\usepackage{comment}
\usepackage{amsmath,amssymb}

\begin{document}


\title{Reassessment of the Null Result of the HST Search for Planets in 47 Tucanae}

%% Use \author, \affil, plus the \and command to format author and affiliation 
%% information.  If done correctly the peer review system will be able to
%% automatically put the author and affiliation information from the manuscript
%% and save the corresponding author the trouble of entering it by hand.
%%
%% The \affil should be used to document primary affiliations and the
%% \altaffil should be used for secondary affiliations, titles, or email.

%% Authors with the same affiliation can be grouped in a single
\author{Kento Masuda\altaffilmark{1,2} and Joshua N. Winn\altaffilmark{1}}
%\affil{Department of Astrophysical Sciences, Princeton University, Princeton, NJ 08544, USA}


\altaffiltext{1}{Department of Astrophysical Sciences, Princeton University,
Princeton, NJ 08544, USA}
\altaffiltext{2}{NASA Sagan Fellow}
\email{kmasuda@astro.princeton.edu}

\begin{abstract}

  We revisit the null result of the {\it Hubble Space Telescope}
  search for transiting planets in the globular cluster 47~Tucanae, in
  the light of improved knowledge of planet occurrence from the {\it
    Kepler} mission.  Gilliland and co-workers expected to find 17
  planets, assuming the 47~Tuc stars have close-in giant planets with
  the same characteristics and occurrence rate as those of the nearby
  stars that had been surveyed up until 1999.  We update this result
  by assuming that 47~Tuc and {\it Kepler} stars have identical planet
  populations.  The revised number of expected detections is
  $4.0^{+1.7}_{-1.4}$.  When we restrict the {\it Kepler} stars to the
  same range of masses as the stars that were searched in 47~Tuc, the
  number of expected detections is reduced to $2.2^{+1.6}_{-1.1}$.
  Thus, the null result of the {\it HST} search is less statistically
  significant than it originally seemed.  We cannot reject even the
  extreme hypothesis that 47 Tuc and {\it Kepler} stars have the same
  planet populations, with more than 2-3$\sigma$ significance.  More
  sensitive searches are needed to allow comparisons between the
  planet populations of globular clusters and field stars.
  
\end{abstract}




\section{Introduction} \label{sec:intro}

A milestone in the history of exoplanet detection was the {\it Hubble
  Space Telescope} survey for transiting planets in the globular
cluster 47 Tucanae (NGC~104) by \citet{2000ApJ...545L..47G}.  This was
the first space-based planet survey, as well as the first exploration
of the planet population within globular clusters.\footnote{Apart from
  the curious case of the candidate planet orbiting a
  pulsar/white-dwarf binary in M4 \citep{1993Natur.365..817B}.}
Despite observing $\approx$34,000 stars nearly continuously for 8.3
days, with a precision high enough to detect giant planets, the
authors did not find any planets. They concluded that hot Jupiters
(HJs) in 47~Tuc are rarer by at least an order of magnitude than in
the solar neighborhood.  Based on inject-and-recover tests and
assumptions about planet occurrence that were reasonable at the time,
they should have detected about 17 planets if the stars in 47~Tuc and
field stars had HJs with the same prevalence.

Over time, this result has come to be regarded as unsurprising. There
are many reasons to expect HJ occurrence in globular clusters to be
lower than in nearby stellar populations, the most obvious reason
being metallicity.  In the local neighborhood, the occurrence of
short-period giant-planet occurrence is strongly associated with high
metallicity \citep{2001A&A...373.1019S, 2005ApJ...622.1102F}, and 47
Tuc has a low metallicity of $-0.7$ \citep{2008ApJ...684..326M}. Other
reasons have also been given. For example, giant planet formation or
migration may be inhibited in environments with radiation from nearby
massive stars \citep{2000A&A...362..968A, 2004ApJ...611..360A,
  2013MNRAS.431...63T}. Planets in globular clusters may be lost
during stellar encounters \citep{1992ApJ...399L..95S,
  2001MNRAS.324..612D, 2001MNRAS.322..859B, 2006ApJ...640.1086F,
  2009ApJ...697..458S}. The clusters are old enough that HJs may have
undergone tidal orbital decay \citep{2010ApJ...723.1703D} or
Roche-lobe overflow due to tidal heating and expansion
\citep{2003ApJ...588..509G}.

While these reasons may seem compelling, they are not necessarily
correct. The cause/effect relationship between metallicity and hot
Jupiters has not been demonstrated. It is conceivable that metallicity
{\it per se} is irrelevant, and that other factors are important which
are associated with high metallicity in the local neighborhood but not
in 47 Tuc. Likewise, it is difficult to anticipate all the
consequences of stellar encounters. Surely they disrupt some planetary
systems, but they might also enhance the rate of HJ production through
high-eccentricity migration. And if HJs form {\it in situ} in
tight orbits \citep{2016ApJ...829..114B}, encounters might be
irrelevant. There may even be modes of HJ formation specific
to globular clusters. In short, since neither HJ formation
nor globular cluster formation are understood, we should perform
observational tests of even the most seemingly obvious expectations.

At the time of this pioneering {\it HST} survey, only one transiting
HJ was known: HD~209458b
\citep{2000ApJ...529L..45C,2000ApJ...529L..41H}. Naturally, in
interpreting their null result, \citet{2000ApJ...545L..47G} assumed
that all HJs resemble this particular planet, which was drawn from
Doppler surveys of nearby stars.  After more than 15 years we have a
better grasp on the prevalence and radius/period distribution of giant
planets, which are crucial for evaluating the number of expected
detections in a transit survey.  We decided to test whether the
conclusions of \citet{2000ApJ...545L..47G} are still valid.  Data from
the most recent space-based transit survey, the NASA {\it Kepler}
mission \citep{2010Sci...327..977B}, are the best available for this
purpose.  In this work, we recalculate the number of expected
detections in the 47~Tuc survey, based on the planet statistics from
{\it Kepler}.

%%%%%%%%%%%%%%%%%%%%%%%%%%%%%%%%%%%%%%%%%%%%%%
% Method
%%%%%%%%%%%%%%%%%%%%%%%%%%%%%%%%%%%%%%%%%%%%%%
\section{Method: Direct Sampling from the {\it Kepler} Transiting Planets} \label{sec:method}

Rather than relying on planet occurrence rates and distributions that
have been inferred from the {\it Kepler} data, we adopt a more direct
procedure.  First we construct a set $\mathcal{S}$ of {\it Kepler}
stars with the same number of members as the sample of 47~Tuc stars
searched by \citet{2000ApJ...545L..47G}. We do so by randomly choosing
entries from the {\it Kepler} Input Catalog (KIC). We thereby
associate each star in 47~Tuc with a {\it Kepler} star. If the {\it
  Kepler} star has detected transiting planets, we assume that the
corresponding star in 47~Tuc has planets with the same properties.
Then we count the number of transiting planets that should have been
detected by \citet{2000ApJ...545L..47G}, taking into account the
sensitivity of their detection pipeline and the (mild) differences in
the transit probabilities between 47~Tuc stars and the {\it Kepler}
stars. This whole procedure is repeated many times, to derive the
probability distribution for the number of expected detections.

For each realization of $\mathcal{S}$ we compute the number of
expected detections,
\begin{equation}
	\label{eq:nobs}
	n_{\rm det} 
	= \sum_{i\,\in\,\mathcal{S}} 
	c_i \cdot n_{{\rm det},i}.
\end{equation}
Here, $n_{{\rm det},i}$ is the number of transiting planets that would
have been detected around the $i$th star in $\mathcal{S}$, assuming it
has planets with the same properties as the associated {\it Kepler}
star. In most cases, of course, the {\it Kepler} star does not have
any detected transiting planets, and $n_{{\rm det},i}=0$.  The
dimensionless factor $c_i$ accounts for the difference in transit
probability between the 47~Tuc star and the associated {\it Kepler}
star (see Section~\ref{ssec:method_geometry}).

To obtain $n_{{\rm det},i}$, we calculate the product of the
detectability $d$ of each planet around that star and the probability
$(1-{\rm FPP})$ that the planet is not a false positive, summed over
the set $\mathcal{P}_i$ of all the transiting planets around that
star.  The detectability $d$ depends on the planet's radius $r$ 
and orbital period $P$ as well as the star's apparent magnitude $V$ (see
Section~\ref{ssec:method_efficiency}). Thus:
\begin{equation}
	\label{eq:nplanet}
	n_{{\rm det},i} =
	\sum_{j\,\in\,\mathcal{P}_i}
	(1 - {\rm FPP}_j) \cdot d(r_j, P_j, V_i).
\end{equation}

The following subsections describe this
calculation in more detail.

\subsection{Sources of Data} \label{ssec:method_data}

For the parameters of the 47~Tuc stars, we use a list of the $V$
magnitudes for the stars searched with {\it HST} that was kindly
provided by R.\ Gilliland.  We adopt the list of {\it Kepler} target stars 
and their planet properties from Data Release (DR)~24, which includes the most recent
catalog of planets and planet candidates.  To assign masses and radii
to the {\it Kepler} stars, we use the posterior probability
distributions from the DR~25 catalog
\citep{2016arXiv160904128M}.\footnote{Data downloaded from
  \url{http://exoplanetarchive.ipac.caltech.edu/bulk_data_download/}.}

\subsection{Simulated Star Sample: $\mathcal{S}$} \label{ssec:method_stars}

For each roll of the dice in our Monte Carlo procedure, we perform
the following steps:
\begin{enumerate}
\item Construct a sample $\mathcal{S}_{\rm K}$ of {\it Kepler} stars for
  which the {\it Kepler} planet catalog is complete for the types of
  planets that could have been found in 47~Tuc.
\item Construct a sample $\mathcal{S}$ of $34,091$ main-sequence stars in 47~Tuc
  and their relevant properties.
\item Associate each star in $\mathcal{S}$ with a star drawn
  randomly from $\mathcal{S}_{\rm K}$.
\end{enumerate}
  
For step 1, each {\it Kepler} star is assigned a mass and radius by
drawing randomly from the posterior distributions for those
quantities. Then we identify the subset of those stars for which a
planet with radius $0.5~R_{\rm Jup}$ and period 8.3~days would have
been detected with a multiple-event-statistic (MES) of 17. The MES is
computed as
\begin{equation}
	\mathrm{MES}=\sqrt{\frac{T_{\rm obs}}{8.3\,\mathrm{days}}}\,
	\frac{(0.5R_{\rm Jup}/R_\star)^2}{\sigma_{\rm CDPP}(T)},
\end{equation}
where $T_{\rm obs}$ is the product of the data span and duty cycle,
and $\sigma_{\rm CDPP}(T)$ is the robust root-mean-squared Combined
Differential Photometric Precision for the timescale of the
corresponding transit duration
\citep{2010arXiv1001.2010W},
\begin{equation}
	\label{eq:duration}
	T=13\,\mathrm{hr} \left(\frac{8.3\,\mathrm{days}}{1\,\mathrm{yr}}\right)^{1/3}
	\left(\frac{\rho_\star}{\rho_\odot}\right)^{-1/3}\times\frac{\pi}{4}.
\end{equation}
Here the mean stellar density $\rho_\star$ is computed from the mass
and radius assigned as above, and the last factor $\pi/4$ comes from
averaging over the impact parameter. The CDPP has only been tabulated
for certain timescales between 1.5 and 15~hours.  When $T$ is within
that range, we compute $\sigma_{\rm CDPP}(T)$ by linear interpolation;
otherwise, following \citet{2015ApJ...809....8B}, we adopt the value
that is tabulated for the closest available timescale.  We also
exclude the stars with $T_{\rm obs}<3\times8.3\,\mathrm{days}$ because
three transits are required for detection. A small number of stars for
which CDPP values are unavailable are also omitted. These criteria
typically leave us with about $172,000$ {\it Kepler} stars in
$\mathcal{S}_{\rm K}$. The fluctuations in the size of
$\mathcal{S}_{\rm K}$ arising from the random sampling of masses and
radii are of order 0.1\%.  The size of $\mathcal{S}_{\rm K}$ is also
insensitive to the exact choice of MES threshold; when we lower it
from 17 to 10, the number of stars increases by less than $1\%$.

\begin{figure}
	\centering
	\includegraphics[width=\columnwidth]{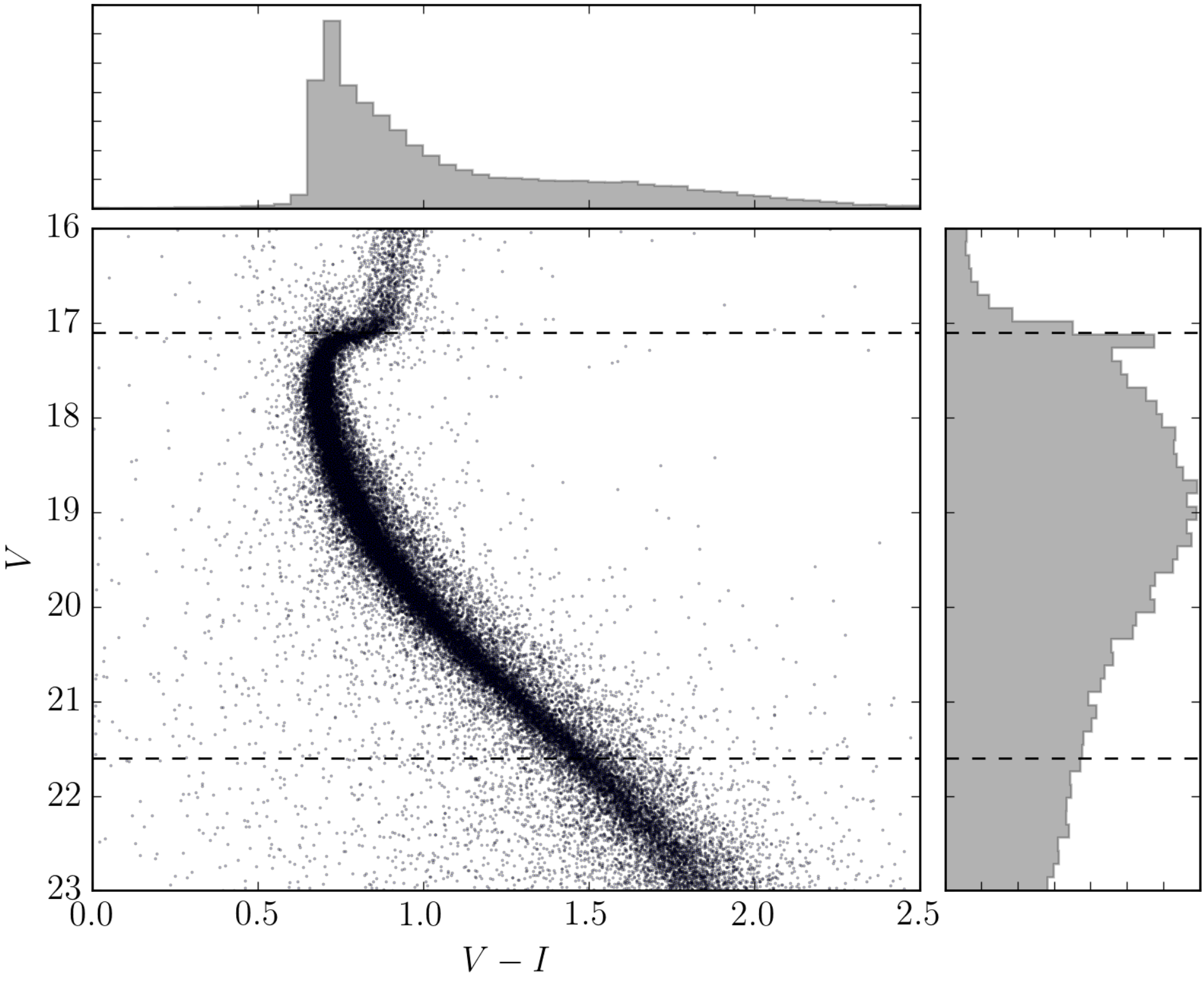}
	\caption{Color-magnitude diagram of 47~Tuc, based on data
          from \citet{2000ApJ...545L..47G}. Just as in that study, we
          select stars between $V=21.6$ and $17.1$ (dashed
          lines).\label{fig:47tuc_hr} }
\end{figure}

For step 2, we draw $34,091$ stars by randomly sampling from the
$35,101$ stars in 47~Tuc that satisfy $17.1<V<21.6$, the same
criterion used by \citet{2000ApJ...545L..47G} to select main-sequence
stars. Figure~\ref{fig:47tuc_hr} shows the color-magnitude diagram for
47~Tuc. The reason why only $34,091$ (and not $35,101$) stars were
searched for planets is that \citet{2000ApJ...545L..47G} imposed a
secondary selection based on $V-I$; see their Figure~1. We did not
impose the same $V-I$ criterion because its functional form is not
readily available. The resulting differences between our simulated
samples, and the actual sample searched by
\citet{2000ApJ...545L..47G}, are very minor and negligible for our
purpose. Each star in $\mathcal{S}$ is assigned a mass and radius
based on its $V$ magnitude and the stellar-evolutionary models of
\citet{1992ApJS...81..163B}, matching the procedure of
\citet{2000ApJ...545L..47G}.

In Step 3, we construct a sample $\mathcal{\tilde S}_{\rm K}$ by
randomly resampling (with replacement) the same number of stars from
$\mathcal{S}_{\rm K}$.  This allows us to take into account the
Poisson fluctuations in the occurrence rate of planets in the {\it
  Kepler} sample, although this source of uncertainty turns out to be
minor.  Then, each star in $\mathcal{S}$ is associated with a star in
$\mathcal{\tilde S}_{\rm K}$ by randomly drawing an entry from
$\mathcal{\tilde S}_{\rm K}$.

%% 47 Tuc CMD

\subsection{Simulated Planet Samples: $\{\mathcal{P}_i\}_{i\,\in\,\mathcal{S}}$ and $\mathrm{FPP}$} \label{ssec:method_planets}

For most stars in $\mathcal{S}$, the corresponding {\it Kepler} star
has no detected transiting planets. In such cases the star in
$\mathcal{S}$ is not assigned any planets.  For cases in which the
{\it Kepler} star does have planets, the corresponding star in
$\mathcal{S}$ is assigned planets with the same orbital period $P$ and
radius $r$.  By ``planets'' we mean KOIs with $P = 0.5$-8.3~days, $r =
0.5$-2~$R_{\rm Jup}$, and a DR24 disposition of either ``confirmed''
or ``candidate''. Because the stellar radii were assigned randomly
from the posterior distribution, the planetary radius $r$ is
recalculated in each realization as the product of the stellar radius
and the planet-to-star radius ratio listed in the KOI catalog.
We neglect the uncertainty in the radius ratio because, for HJs, 
the leading source of uncertainty in $r$ is the uncertainty in the stellar radius.

We also need to compute the factor $(1-{\rm FPP})$ in
Eqn.~\ref{eq:nplanet}, to account for false positives.  We assume FPP~$=0$
for ``confirmed'' KOIs.  For the others, we use the FPPs computed by
\citet{2016ApJ...822...86M}.

\subsection{Correction for Transit Probability: $c$} \label{ssec:method_geometry}

The geometric transit probability is $R_\star/a$, which is
proportional to $\rho_\star^{-1/3}$ at fixed orbital period.  Since
each star in $\mathcal{S}$ has a different mean density than the
corresponding KIC star, we need to
correct for the difference in transit probability.
We do so by modifying the planet count
around the $i$th star, $n_{{\rm det},i}$, by the factor
\begin{equation}
  c = \left( \frac{\rho_{\star, \rm K}}
             {\rho_{\star, \rm 47~Tuc}} \right)^{1/3},
\end{equation}
where $\rho_{\star, \rm K}$ and $\rho_{\star, \rm 47~Tuc}$ are the mean densities 
of the {\it Kepler} and 47~Tuc stars associated with the $i$th star, respectively.

\subsection{Detection Efficiency: $d$} \label{ssec:method_efficiency}

\begin{figure}
	\centering
	\includegraphics[width=\columnwidth]{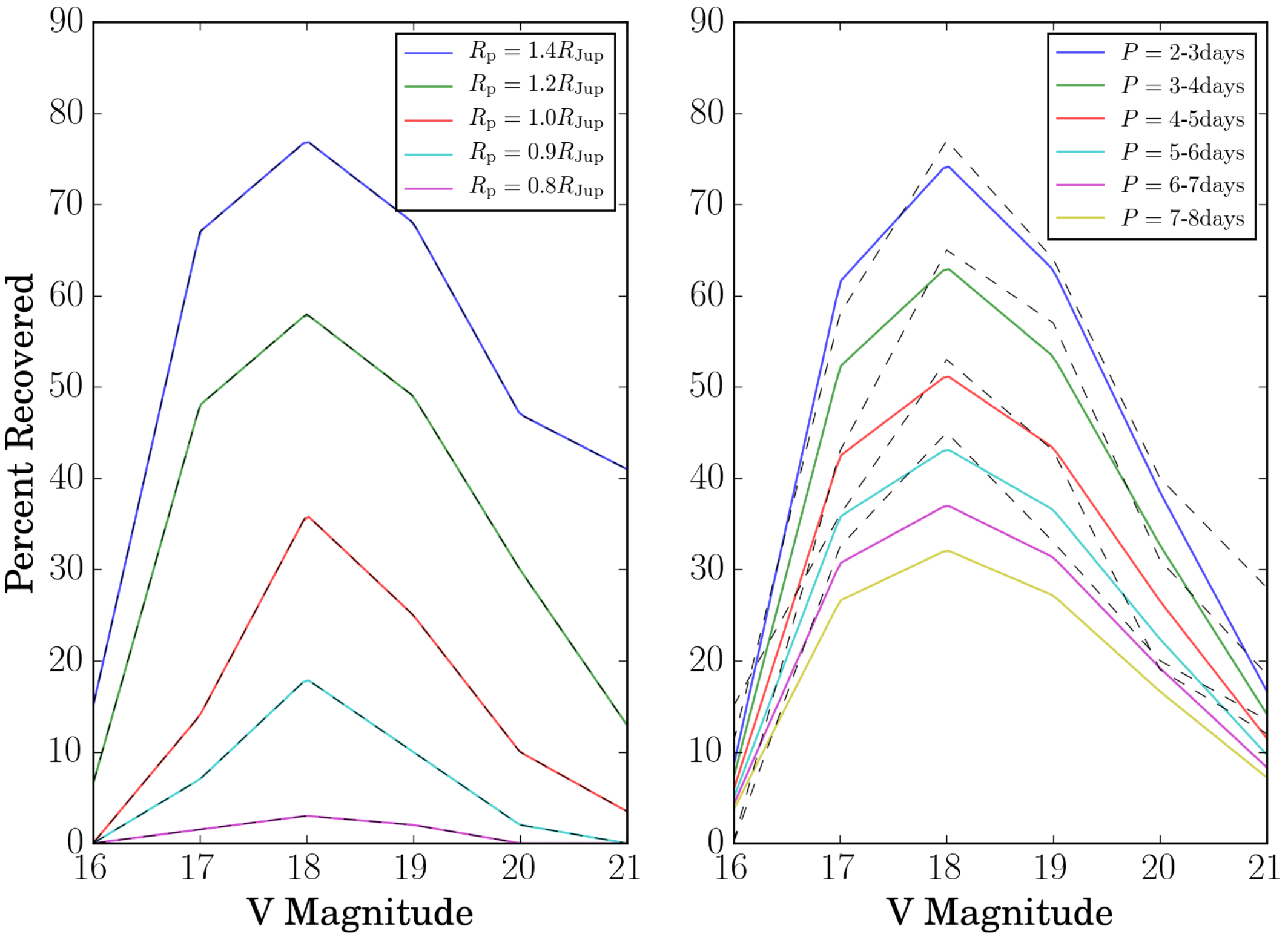}
	\caption{Dependence of planet detectability on planetary
          radius (left) and orbital period (right) as a function of
          apparent $V$ magnitude.  Dashed lines are the numerical
          results of \citet{2000ApJ...545L..47G}.  Solid lines are the
          output of our analytic fitting function
          (Eqn.~\ref{eq:d_analytic}).  The numerical and analytic
          results are coincident in the left panel, by construction.
          In the right panel the agreement is good to within
          about 10\%. \label{fig:d_analytic} }
\end{figure}

\citet{2000ApJ...545L..47G} used inject-and-recover simulations to
determine the detection efficiency of their transit search as a
function of $r$, $P$, and $V$. The results are presented graphically
in their Figure~4, and are reproduced in our
Figure~\ref{fig:d_analytic} along with an analytic fitting function
we constructed to match the numerical results.
The fitting function is of the form
\begin{equation}
	\label{eq:d_analytic}
	d(r, P, V) = f(r, V)\,\frac{g(P)}{\langle g\rangle}.
\end{equation}
The function $f$ is computed by linear interpolation of the data
presented in the left panel of Figure~4 of \citet{2000ApJ...545L..47G}.  In some cases
we need to extrapolate $f$ beyond the ranges plotted by
\citet{2000ApJ...545L..47G}: we assume $f=0$ for $r\leq0.6\,R_{\rm
  Jup}$; $f$ achieves its maximum value at $r=1.4\,R_{\rm Jup}$; and
$f(r, 22)=f(r, 21)$. The function
\begin{equation}
	g(P) = 0.77\exp[-0.27(P-2.95)^{0.75}]
\end{equation}
is designed to match the right panel of the same figure.
It matches the $P$-dependence at $V=18$ and $r=1.2\,R_{\rm Jup}$,
achieves its maximum value for $P\lesssim2.5\,\mathrm{days}$, 
and allows for continuous extrapolation to $P=8.3\,\mathrm{days}$.
The normalization $\langle g \rangle=0.6$ represents the average
over $P=2\mathchar`-6\,\mathrm{days}$, such that $f=d$ when
averaging over this period range.

%% analytic detectability

%%%%%%%%%%%%%%%%%%%%%%%%%%%%%%%%%%%%%%%%%%%%%%%%
%% Results
%%%%%%%%%%%%%%%%%%%%%%%%%%%%%%%%%%%%%%%%%%%%%%%%
\section{Results} \label{sec:results}

\begin{figure*}
	\centering
	\includegraphics[width=1.6\columnwidth]{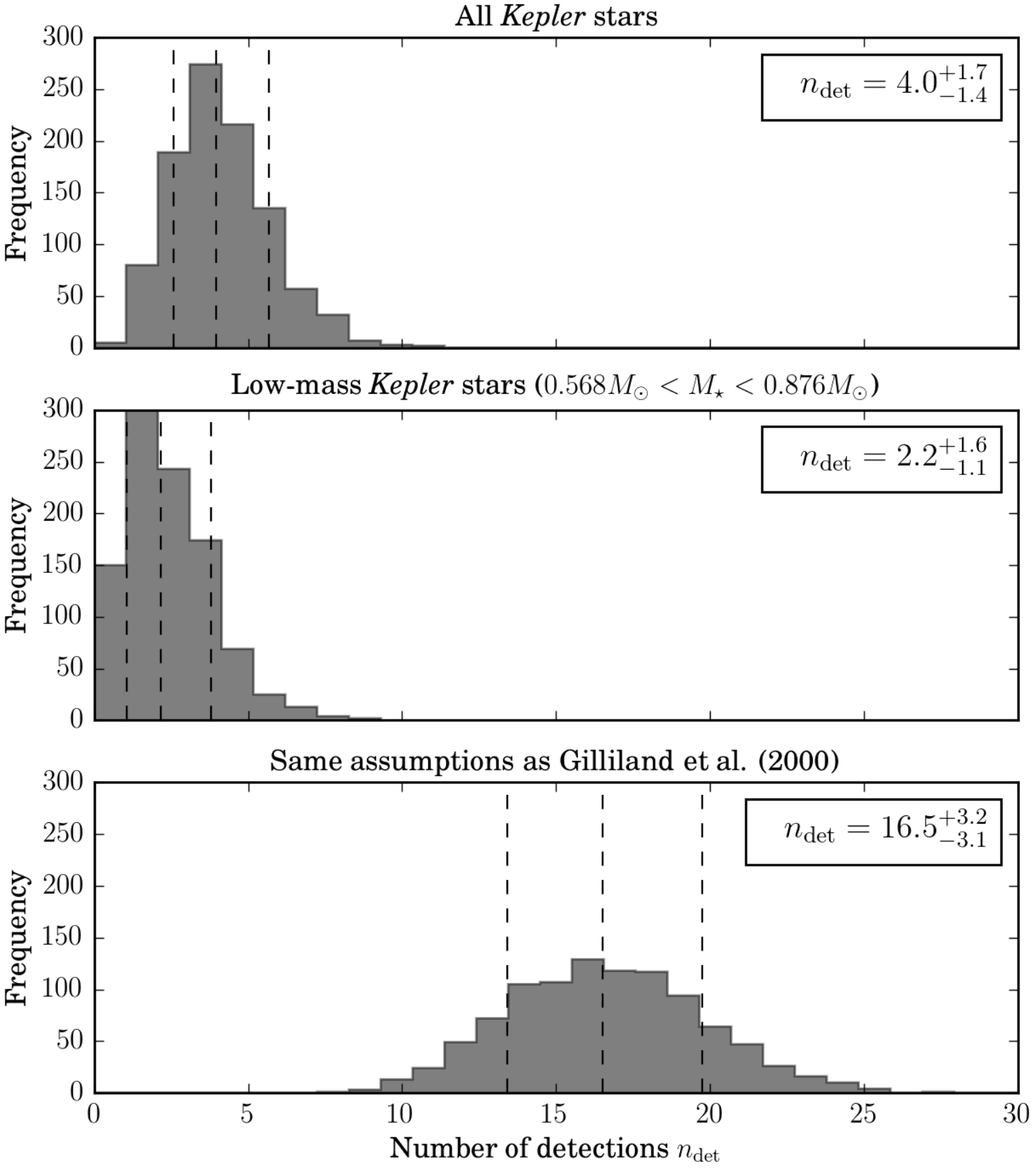}
	\caption{Distributions of the expected number of detections
          $n_{\rm det}$ from $1000$ simulations. From top to bottom,
          the results based on all the {\it Kepler} stars, 
          mass-controlled subset of the {\it Kepler} stars, and
          the same assumptions as adopted in
          \citet{2000ApJ...545L..47G} are shown. The
          vertical dashed lines and the numbers in the upper right
          boxes show $15.87\%$, $50\%$, and $84.13\%$ percentiles of
          the distributions.
	 \label{fig:hist}
	}
\end{figure*}

We construct 1000 realizations of the star and planet samples,
following the procedures described in the previous section.  In each
case we compute $n_{\rm det}$, the number of transiting planets that would have been
detected in the {\it HST} search of 47~Tuc.  The top panel of
Figure~\ref{fig:hist} shows the result: the expected number of
detections is $4.0^{+1.7}_{-1.4}$. Here and elsewhere the quoted value
is the median of the probability distribution of $n_{\rm det}$, and the
uncertainty interval covers 68.3\% of the probability surrounding the
median (a ``one-sigma'' interval).

We also constructed another 1000 realizations, this time restricting
the masses of the {\it Kepler} stars to the range
0.568-0.876~$M_\odot$, the same range of masses as the 47~Tuc stars
satisfying $17.1<V<21.6$.
We perform this test because there is evidence that the planet
population around low-mass stars differs from that of high-mass stars.
In this case, $\mathcal{S}_{\rm K}$ consists of approximately $58,000$
stars, and the expected number of detections is $2.2^{+1.6}_{-1.1}$.
This result is shown in the middle panel of Figure~\ref{fig:hist}.

As a check on our procedure, we constructed an additional 1000
realizations of $\mathcal{S}$, this time assigning planets based on
the same assumptions as \citet{2000ApJ...545L..47G} instead of using
{\it Kepler} data.  Specifically, we assume that HJs exist around
0.9\% of all stars, with a transit probability of $10\%$, and that all
HJs have $r=1.3~R_{\rm Jup}$, and $P=3.5\,\mathrm{days}$.  The radius
and period are those of HD~209458b, the only HJ that was known at the
time. In this case we find $n_{\rm det} = 16.5^{+3.2}_{-3.1}$, as
shown in the bottom panel of Figure~\ref{fig:hist}. This agrees with
the conclusions of \citet{2000ApJ...545L..47G}, validating our process
for constructing $\mathcal{S}$ and simulating the detection
efficiency.

\begin{figure*}
	\centering
	\includegraphics[width=1.6\columnwidth]{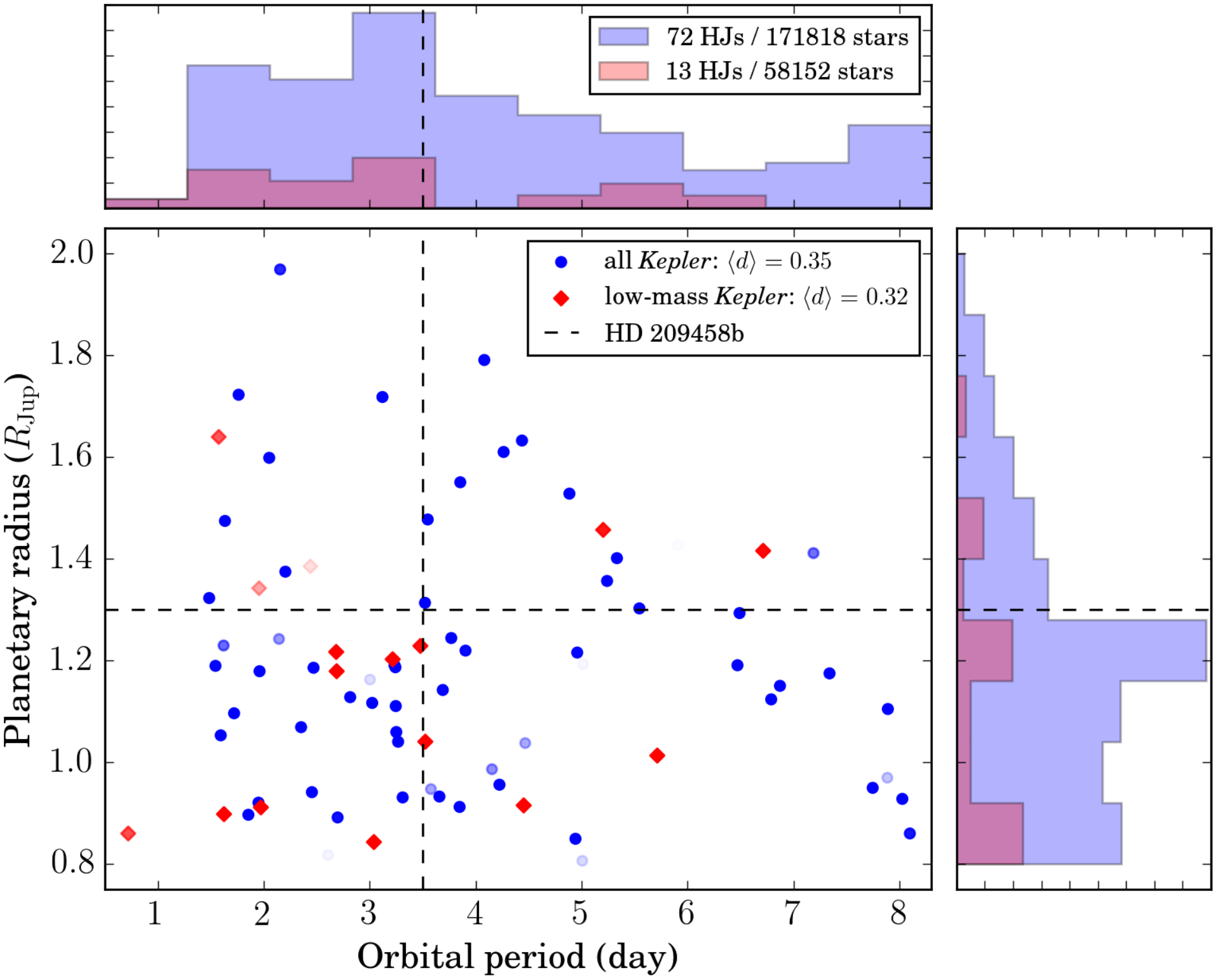}
	\caption{Planetary radius--orbital period distribution of the
          HJs around {\it Kepler} stars in one realization of
          $\mathcal{S}_{\rm K}$ taken from the simulations. Here we
          only show the planets larger than $0.8R_{\rm Jup}$, i.e.,
          the ones that are essentially detectable. Blue filled
          circles are HJs around all the {\it Kepler} stars, while red
          filled diamonds show those around the low-mass subset
          ($0.568\,M_\odot<M_\star<0.876\,M_\odot$). Opacity of the points reflect the
          FPP values of each planet (planets with higher FPPs are more
          transparent). Vertical and horizontal dashed lines indicate
          the values of HD 209458b ($P=3.5\,\mathrm{days}$ and $r=1.3\,R_{\rm Jup}$), 
          which are assumed by \citet{2000ApJ...545L..47G}. 
          The inset in the top panel
          shows the sum of $1-\mathrm{FPP}$ for all the plotted HJs and
          the number of KIC stars in this $\mathcal{S}_{\rm K}$. The
          inset in the middle panel shows detection efficiency $d$
          averaged over all the HJs in the plot with the weight
          $1-\mathrm{FPP}$. \label{fig:koi_hj_prop} }
\end{figure*}

The {\it Kepler}-based simulations give a smaller number of expected
detections than 17. Table~\ref{tab:analytic} breaks down the reasons
for the difference.  One factor is the lower occurrence rate of HJs in
the {\it Kepler} field compared to the value assumed by
\citet{2000ApJ...545L..47G}.  The {\it Kepler} occurrence rate is even
lower when we restrict the range of stellar masses to
0.568-0.876~$M_\odot$. Another important factor is that the typical
value of detectability in our planet sample is only about $60\%$ of
the value for the HJ assumed in \citet{2000ApJ...545L..47G} (second
row in Table \ref{tab:analytic}).  This is because the detectability
is a strong function of planet radius, and the HJs in the {\it Kepler}
field are often smaller than $1.3\,R_{\rm Jup}$. This situation is
illustrated in Figure \ref{fig:koi_hj_prop}, which shows the $r$-$P$
distribution of the HJs around the {\it Kepler} stars for one
realization of $\mathcal{S}_{\rm K}$.

%% Results

%% HJ property

%%%%%%%%%%%%%%%%%%%%%%%%%%%%%%%%%%%%%%%%%%%%%%%%
%% Table
%% Analytic interpretation
%%%%%%%%%%%%%%%%%%%%%%%%%%%%%%%%%%%%%%%%%%%%%%%%
\begin{deluxetable*}{lcccc}
\tablecaption{Typical Values of Each Factor in Equations (\ref{eq:nobs}) and (\ref{eq:nplanet}) from our Simulations\label{tab:analytic}}
\tablehead{
\colhead{} & \colhead{\citet{2000ApJ...545L..47G}} & \colhead{All {\it Kepler}}
& \colhead{Low-mass {\it Kepler}} & \colhead{RV Sample}
}
\startdata
Transiting HJ Occurrence $\sum_\mathcal{P} (1-\mathrm{FPP})/34,091$ & $(8\mathchar`-10)\times10^{-4}$ & $3.9\times10^{-4}$ & $1.8\times10^{-4}$ & $14\times10^{-4}$\\
Average Detectability $\langle d \rangle$	& $0.54$  & $0.34$ & $0.31$  & $0.38$\\
Average Transit-Probability Correction $\langle c \rangle$ & \nodata & $0.84$ & $1.1$ & $0.79$\\
Yield from $34,091$ Stars & $17$ & $4.0$  & $2.2$  & $15$ \\ \enddata
\tablecomments{In columns 2 and 3, the values of transiting HJ
  occurrence represent $1-\mathrm{FPP}$ summed over practically detectable HJs
  in $\mathcal{P}$ ($P=0.5$-$8.3$~days, $r=0.8$-2~$R_{\rm Jup}$),
  divided by $34,091$. The values for detectability and the
  transit-probability correction are the averages for all the HJs in
  this range, weighted by $1-\mathrm{FPP}$. The quoted values are the
  medians based on 1000 simulations.  Note that the last row is
  approximately the product of the first three rows and $34,091$.  }

\end{deluxetable*}

%%%%%%%%%%%%%%%%%%%%%%%%%%%%%%%%%%%%%%%%%%%%%%%%
%% Discussions
%%%%%%%%%%%%%%%%%%%%%%%%%%%%%%%%%%%%%%%%%%%%%%%%
\section{Discussion} \label{sec:discussion}

\subsection{Occurrence Rate of {\it Kepler} Hot Jupiters} \label{ssec:discussion_occurrence}

One of the critical factors that reduce the number of expected
detections is the lower occurrence rate of transiting HJs in the {\it
  Kepler} sample, compared to the rate assumed by
\citet{2000ApJ...545L..47G}.  \citet{2012ApJS..201...15H} also
measured the HJ occurrence rate based on {\it Kepler} data.  To
facilitate a comparison between their study and ours, we calculate the
occurrence rate of HJs, as opposed to transiting HJs. To do so we
perform another set of simulations, this time with
$\mathcal{S}=\mathcal{\tilde S}_{\rm K}$, and replacing
Eqn.~\ref{eq:nplanet} with
\begin{equation}
	\label{eq:nplanet-hj}
	n_{{\rm det},i} =
	\sum_{j\,\in\,\mathcal{P}_i}
	(1 - {\rm FPP}_j) \frac{a_j}{R_{\star,i}},
\end{equation}
i.e., we divide by the transit probability $R_\star/a$.  We adopt the
value of $a/R_\star$ from the KOI catalog, assuming a circular orbit
and neglecting the uncertainty. For consistency with
\citet{2012ApJS..201...15H} we also modify the definition of
``planets" in Section~\ref{ssec:method_planets} to be those with
$P<10\,\mathrm{days}$ and $r = 0.8$-2~$R_{\rm Jup}$.  Through this
procedure we find $f_{\rm HJ}=0.43^{+0.07}_{-0.06}\%$, in agreement
with the value of 0.4\% found by \citet{2012ApJS..201...15H}.

When restricting the {\it Kepler} stars to the same range of masses as
the stars that were searched in 47~Tuc, we find $f_{\rm
  HJ}=0.24^{+0.10}_{-0.09}\%$, which is smaller than the rate obtained
for the entire sample of stars. This suggests that the low-mass {\it
  Kepler} stars have an even lower HJ occurrence rate. The statistical
significance of the difference is modest, because of the relatively
small number of HJs ($\approx$10) in the restricted sample.  This
possible dependence of HJ occurrence on stellar mass will come into
sharper focus after the {\it TESS} mission
\citep{2014SPIE.9143E..20R}, which should provide a larger sample of
transiting HJs around a wide range of stellar types.

\subsection{Comparison with RV Samples} \label{ssec:discussion_rv}

\citet{2012ApJ...753..160W} measured the HJ occurrence rate using the
Doppler or radial-velocity (RV) technique. Based on a sample of 10 HJs
found within a set of 836 stars, they found $f_{\rm HJ} = 10/836 =
(1.2\pm 0.4)\%$, which is higher than our result by 1.9$\sigma$.  If
the occurrence rate is really 1.2\%, the number of expected detections
in the 47~Tuc survey would be much higher than the results presented
in the previous section.  To demonstrate this, we perform another
round of simulations using the RV sample from
\citet{2012ApJ...753..160W} instead of {\it Kepler} stars;
this time $\mathcal{S}_{\rm K}$ consists of $836$ stars considered in the RV sample
of \citet{2012ApJ...753..160W}, among which $10$ are associated with HJs
listed in Table~\ref{tab:rvsample}.
One obstacle is that most of the HJs in the RV sample do not transit, and their radii are unknown;
even their true masses are unknown.  We must nevertheless assign them
radii in our simulations. We do so by assigning each planet a random
orbital orientation (uniform in $\cos I$) and calculating the planet
mass $m$ based on the measured value of $m\sin I$.  Then we calculate
$r$ using the relations between planetary mass, radius, and incident
flux $F$ presented by \citet{2013ApJ...768...14W}. We also add Gaussian
random deviates to $r$ to account for the scatter in the $m/r/F$
relation ($1.15\,R_\oplus$ for $m>150\,M_\oplus$ and $1.41\,R_\oplus$
for smaller $m$).  The result for the number of expected detections in
the 47~Tuc survey is $15.2^{+7.1}_{-5.9}$.
This is larger than our {\it Kepler}-based results and compatible with
the original estimate of \citet{2000ApJ...545L..47G}.  The difference
is mainly due to the higher $f_{\rm HJ}$ of the RV sample, with a
smaller contribution from somewhat higher detectability (larger planets).

The RV-based result has a higher statistical uncertainty than our {\it
  Kepler}-based result. There are a few additional reasons to attach
greater weight to the {\it Kepler}-based result. The RV sample was
constructed {\it post facto} from stars originally selected for
undocumented reasons. \citet{2011arXiv1109.2497M} performed an
independent RV-based analysis of similar stars, finding 5~HJs within a
sample of 822 stars (see their Sec.~4.2), and giving $f_{\rm HJ} =
0.6_{-0.2}^{+0.3}\%$.  This is half the value reported by
\citet{2012ApJ...753..160W}, and within 1$\sigma$ of the {\it
  Kepler}-based result. Table 1 of \citet{2011arXiv1109.2497M} reports
a higher rate of 0.89\%, but this includes planets with masses as low
as 0.16~$M_{\rm Jup}$ and periods $<$11~days rather than 10~days. We
do not know why the seemingly more arbitrary upper limit of 11 days
was chosen, illustrating the difficulty of analyzing {\it post facto}
samples. Furthermore, our method is more direct by associating real
planets and their properties to 47~Tuc stars, rather than inferring an
occurrence rate for a certain sharply-defined category of planets from
one survey, and then using that rate to interpret the results from a
different survey.

A separate issue is that the RV surveys do not provide much
information about the range of stellar masses (0.568-0.876~$M_\odot$)
spanned by the 47~Tuc stars. The RV sample of
\citet{2012ApJ...753..160W} includes only one HJ in that mass range,
causing a large Poisson uncertainty in the occurrence rate.

%%%%%%%%%%%%%%%%%%%%%%%%%%%%%%%%%%%%%%%%%%%%%%%%
%% Table
%% RV Sample
%%%%%%%%%%%%%%%%%%%%%%%%%%%%%%%%%%%%%%%%%%%%%%%%
\begin{deluxetable*}{cccccccc}
\tablecaption{Properties of HJs and their Host Stars in the RV Sample of \citet{2012ApJ...753..160W}\label{tab:rvsample}}
\tablehead{
\colhead{Name} & \colhead{$m\sin I$ ($M_{\rm Jup}$)} & \colhead{$P$ (day)}
& \colhead{$M_\star$ ($M_\odot$)} & \colhead{$R_\star$ ($R_\odot$)} & \colhead{$T_{\rm eff}$ (K)} & \colhead{[Fe/H]} &  \colhead{Reference\tablenotemark{a}}
}
%\colnumbers
\startdata
$\upsilon$ And (HD 9826) b & $0.669\pm0.026$ & $4.6$ & $1.31\pm0.03$ & $1.573\pm0.019$ & $6213$ & $0.15$ & 1\\
$\tau$ Boo (HD 120136) b	& $4.12\pm0.15$	& $3.3$ & $1.35\pm0.03$ & $1.419\pm0.019$ & $6387$ & $0.23$ & 1\\
$51$ Peg (HD 217014) b	& $0.461\pm0.016$ & $4.2$ & $1.10\pm0.03$ & $1.138\pm0.016$ & $5787$ & $0.20$ & 1\\
HD 217107 b	& $1.401\pm0.048$ & $7.1$ & $1.108\pm0.043$ & $1.500\pm0.030$ & $5704$ & $0.389\pm0.030$ & 2\\
HD 185269 b	& $0.954\pm0.069$ & $6.8$ & $1.28\pm0.1$ & $1.88\pm0.1$ & $5980$ & $0.11\pm0.05$ & 3\\
HD 209458 b	& $0.689\pm0.024$ & $3.5$ & $1.18\pm0.06$ & $1.203\pm0.061$ & $6092$ & %$0.01$ (VF05)
$0.00\pm0.05$ & 1,4,7\\
HD 189733 b	& $1.140\pm0.056$ & $2.2$ & $0.846^{+0.068}_{-0.049}$ & $0.805\pm0.016$ & $4875$ & $-0.03\pm0.08$ & 4,5,7\\
HD 187123 b	 & $0.510\pm0.017$ & $3.1$ & $1.037\pm0.025$ & $1.143\pm0.039$ & $5815$ & $0.121\pm0.030$ & 2\\
HD 46375 b	& $0.2272\pm0.0091$ & $3.0$ & $0.93\pm0.03$ & $1.003\pm0.039$ & $5285$ & $0.24$ & 1\\
HD 149143 b	& $1.328\pm0.078$ & $4.0$  & $1.21\pm0.1$ & $1.49\pm0.1$ & $5884$ & $0.26\pm0.05$ & 6\\
\enddata
\tablenotetext{a}{1: \citet{2005ApJS..159..141V}, 2: \citet{2015ApJ...800...22F}. 3: \citet{2006ApJ...652.1724J}, 4: \citet{2015MNRAS.447..846B}, 5: \citet{\dekok}, 6: \citet{2006ApJ...637.1094F},
7: \citet{2008ApJ...677.1324T}}
\end{deluxetable*}

\subsection{Other Globular Cluster Surveys} \label{ssec:discussion_others}

We have focused on the survey by \citet{2000ApJ...545L..47G} because
it is the most sensitive survey that has yet been conducted for
planets in globular clusters.  \citet{2005ApJ...620.1043W} used
ground-based observations to perform a search for transiting planets
in a sample of $21,920$ stars in a less crowded region of 47~Tuc.
They expected to find 7 planets if the planet population were
identical to that of field stars, and found none. However, although
they took into account the period-dependence of the selection
function, they do not appear to have taken into account the much
stronger dependence on planet radius.  Moreover, their expectation was
based on a HJ occurrence rate of 0.8\%, larger than the {\it Kepler}
value.  Using the methodology presented in this paper, we expect that
the number of expected detections would also be reduced by about a factor of
4, as was the case with the {\it HST} survey. This would cause the
apparent difference with field stars to be statistically
insignificant. The same argument would apply to the ground-based
survey of $\omega$ Centauri by \citet{2008ApJ...674.1117W}, which was
only sensitive to relatively large HJs ($\gtrsim$1.5~$R_{\rm
  Jup}$).

\citet{2012A&A...541A.144N} conducted an {\it HST} search for
transiting planets among $5,078$ members of NGC~6397. They were
sensitive to giant planets with periods between $0.2$-$14$~days, and
did not detect any planets. They performed a statistical analysis of a
subsample of $2,215$ M-dwarfs and could not rule out the hypothesis
that the cluster stars have the same planet population as field
stars. This is not surprising, given the relatively small number of
stars in the sample.

\subsection{Metallicity Effect} \label{ssec:discussion_metal}

We controlled for stellar mass by restricting the {\it Kepler}
comparison sample to the same range of masses as the stars that were
searched in 47~Tuc.  In addition to mass, the stellar metallicity is
thought to be strongly linked to the HJ occurrence rate \citep[see,
  e.g.,][]{2010PASP..122..905J}.  It has long been known that a low
stellar metallicity is associated with a low occurrence rate of giant
planets with orbital distances $\lesssim$1~AU. However, with the
available data it is impossible to control for metallicity.  The stars
in 47~Tuc have $\mathrm{[Fe/H]}\approx -0.7$, while {\it Kepler} stars
have a mean [Fe/H]~$\approx 0$ \citep{2014ApJ...789L...3D,
  2016arXiv161201616G}. We cannot restrict the {\it Kepler} sample to
low-metallicity stars because reliable metallicities are only
available for a small number of stars, and most likely the {\it
  Kepler} field does not include enough low-metallicity stars for our
resampling procedure to be effective.

Instead, we simply note that the number of expected detections has
been lowered to such a degree that we are unable to say confidently
whether the lack of detected planets could be attributable to the low
metallicity of 47~Tuc. After controlling for stellar mass (but not
metallicity), the number of expected detections is
$2.2^{+1.6}_{-1.1}$, only marginally inconsistent with zero.
Controlling for metallicity would lower the number of expected
detections still further. For example, \citet{2010PASP..122..905J}
found that giant-planet occurrence scales as
$10^{1.2\mathrm{[Fe/H]}}$; if we assume this is also true of stars in
globular clusters, then the mean number of expected detections becomes
less than unity for $\mathrm{[Fe/H]}\approx -0.29$. 
\citet{2014ApJ...790...91S}
argued for an even stronger dependence on metallicity, with
giant-planet occurrence scaling as $10^{2.3\mathrm{[Fe/H]}}$. Using
that relation, the mean number of expected detections becomes less
than unity for $\mathrm{[Fe/H]}\approx -0.15$.

\subsection{Choice of Stellar Models} \label{ssec:discussion_isochrone}

We adopted stellar parameters for the 47~Tuc stars based on the
stellar-evolutionary models of \citet{1992ApJS...81..163B}, following
\citet{2000ApJ...545L..47G}. More recent stellar-evolutionary models
are available. Adopting a different set of stellar models alters the
stellar mass and radius for a given $V$ magnitude. This affects the
correction for transit probability (Section
\ref{ssec:method_geometry}), and the detectability as a function of
$V$, $r$, and $P$, by altering the transit depth and duration.

To check on the sensitivity of our results to the choice of
stellar-evolutionary models, we recompute the relation between
$(M_\star, R_\star)$ and $V$-magnitude using the Dartmouth isochrones
\citep{2008ApJS..178...89D}.\footnote{We use the online tool:
  \url{http://stellar.dartmouth.edu/models/webtools.html}} We assume a
cluster age of $11.6\,\mathrm{Gyr}$, $\mathrm{[Fe/H]}=-0.69$,
$\mathrm{[\alpha/H]}=0.20$, and helium mass fraction of
$0.2525$. These values are nearly the same as those obtained by
\citet{2016ApJ...823...18C} via isochrone fitting to an {\it HST}
infrared color-magnitude diagram, but are slightly modified to match
the color-magnitude diagram in Figure~\ref{fig:47tuc_hr} with the
distance modulus $13.4$ and $E(V-I)=0.05$.

Using this model, we recompute $\rho_\star^{-1/3}$ (relevant to the
transit probability correction) and
$\sqrt{\rho_\star^{-1/3}R_\star^{-2}}$ (relevant to detectability) for
each of the stars in 47 Tuc and compare them to those computed with
the models of \citet{1992ApJS...81..163B}. We find that the
differences are only a few percent, on average, and no larger than
$25\%$ for any choice of $V$. We conclude that the choice of
stellar-evolutionary models does not significantly affect our results.

\subsection{Effect of Extrapolating Detection Efficiency} \label{ssec:discussion_extrapolation}

To cover the full range of periods and sizes of {\it Kepler} HJs,
we needed to extrapolate the numerical results for detection efficiency 
beyond the limits presented in Figure~4 of \citet{2000ApJ...545L..47G}.
Specifically we assumed
\begin{itemize}
\item $f = 0$ for $r\leq 0.6\,R_{\rm Jup}$. This seems a safe assumption
  because $f$ is already very close to zero at $r=0.8\,R_{\rm Jup}$.
\item $g$ saturates at its maximum value for $P\lesssim 2.5\,\mathrm{days}$.
  This too seems a safe assumption, and (given our functional form) is necessary
  to maintain $d\leq 1$ at short periods.
\item $f$ saturates at its maximum value for $r\geq1.4\,R_{\rm Jup}$, regardless of $V$.
\item $g$ between $P=6$-8.3~days is given by smooth extrapolation from $P<6$~days.
\end{itemize}
The validity of the last two assumptions is not so obvious.
It is conceivable that $f$ could increase beyond $r=1.4\,R_{\rm Jup}$;
this would not violate $d\leq1$ as long as $V\neq18$.
It is also conceivable that $g$ drops abruptly as $P$
approaches the total duration (8.3~days) of the time series that was searched.

To check on the sensitivity of our results to these two assumptions,
we perform additional rounds of simulations using extreme forms of
$d$:
\begin{enumerate}
\item $f = f_{\rm max}$ for $r\geq1.6\,R_{\rm Jup}$, regardless of $V$. Here
  $f_{\rm max}$ is the maximum possible value satisfying $d\leq1$, and is equal
  to $f(1.4\,R_{\rm Jup}, 18)$.
\item $g(P)=0$ for $P\geq6\,\mathrm{days}$.
\end{enumerate}
In the first case we
find $n_{\rm det}=4.1^{+1.8}_{-1.4}$ for the full {\it Kepler} sample,
and $n_{\rm det}=2.3^{+1.7}_{-1.2}$ for the low-mass {\it Kepler}
sample.  In the second case, we find $n_{\rm det}=3.7^{+1.7}_{-1.3}$ and
$n_{\rm det}=2.0^{+1.5}_{-1.2}$ for the full and low-mass samples,
respectively. These results show that our conclusions are fairly
insensitive to the manner in which we have extrapolated the detection efficiency.

%%%%%%%%%%%%%%%%%%%%%%%%%%%%%%%%%%%%%%%%%%%%%%%%
%% Conclusion
%%%%%%%%%%%%%%%%%%%%%%%%%%%%%%%%%%%%%%%%%%%%%%%%
\section{Summary and conclusion}

Among the many gifts of the {\it Kepler} mission is a very large
sample of stars that have been exhaustively searched for the types of
transiting planets that could have been detected in the prior {\it
  HST} survey of 47~Tuc by \citet{2000ApJ...545L..47G}. The {\it
  Kepler} survey thereby provides the best and most reliable means to
try and interpret the null result of the 47~Tuc survey. We have used a
resampling technique to test the hypothesis that the {\it Kepler}
stars and the 47~Tuc stars have the same planet population;
specifically, the same occurrence rate and radius/period distribution
for giant planets. Under this hypothesis, we found that the number of
transiting planets that should have been detected in the 47~Tuc survey
is $4.0^{+1.7}_{-1.4}$.  Thus, the hypothesis can only be rejected at
the $\approx$3$\sigma$ level.  We also tested the hypothesis that the
47~Tuc stars and the {\it Kepler} stars {\it over the same range of
  mass} have the same population of close-in giant planets. In this
case we find that only $2.2^{+1.6}_{-1.1}$ planets should have been
detected in the 47~Tuc survey, and there is a $\approx$15\% chance
that no planets would be found.  Both of these results lead to a lower
degree of confidence that the planet populations are different than
was originally thought.

The null result reported by \citet{2000ApJ...545L..47G} remains the
best constraint on the planet occurrence in globular clusters ever
obtained, and suggests that close-in giant planets are rarer in 47~Tuc
than in the field, but with a low statistical significance.  We are
therefore still far from understanding the planet population within
globular clusters, and what might cause it to differ from that of
other types of stars.  A more sensitive search for planets in globular
clusters is needed.

\acknowledgments

We are grateful to Ron Gilliland, for providing us with the stellar
properties of the stars that were searched in 47~Tuc, and for his
comments on the manuscript. We thank the referee, William Cochran, 
for his constructive suggestions. We also thank Luke Bouma and Tim Morton
for helpful discussions. This work was performed under contract with
the California Institute of Technology (Caltech)/Jet Propulsion
Laboratory (JPL) funded by NASA through the Sagan Fellowship Program
executed by the NASA Exoplanet Science Institute.

%% DRAFT END




\bibliographystyle{aasjournal}


\listofchanges

\end{document}